\documentclass{aa}  

\usepackage{graphicx}
\usepackage{txfonts}
\usepackage{newtxtext,newtxmath}

\usepackage[T1]{fontenc}
\usepackage{ae,aecompl}

\usepackage{graphicx}	
\usepackage{amsmath}	
\usepackage{amssymb}	
\usepackage{mathrsfs}
\usepackage{breqn}
\usepackage{placeins}
\usepackage{import}
\usepackage{ulem}
\usepackage{mathtools}
\usepackage[svgnames]{xcolor}
\usepackage[colorlinks = true,
            linkcolor = blue,
            urlcolor  = blue,
            citecolor = blue,
            anchorcolor = blue]{hyperref}

\makeatletter
\newcommand{\setword}[2]{%
  \phantomsection
  \def\@currentlabel{\unexpanded{#1}}\label{#2}%
}
\makeatother
\renewcommand{\vec}[1]{\boldsymbol{#1}}

\newcommand{\DS}{\displaystyle}

\newcommand{\pd}[2]{\frac{\partial #1}{\partial #2}}

\begin{document}

   \title{A numerical study of the interplay between Fermi acceleration mechanisms in radio lobes of FR-II radio galaxies}

   \titlerunning{Fermi acceleration mechanisms in radio lobes of FR-II radio galaxies}

   \author{Sayan Kundu
          \inst{1}\fnmsep\thanks{sayan.astronomy@gmail.com}
          \and
          Bhargav Vaidya
          \inst{1}
          \and
          Andrea Mignone
          \inst{2}
          \and
          Martin J. Hardcastle
          \inst{3}
          }

   \institute{Discipline of Astronomy, Astrophysics           and Space Engineering, Indian  
             Institute of Technology, Indore, Madhya Pradesh, India - 452020
             \and
             Dipartimento di Fisica Generale, Universita degli Studi di Torino, Via Pietro Giuria 1, 10125 Torino, Italy
             \and
             Centre for Astrophysics Research, Department of Physics, Astronomy and Mathematics, University of Hertfordshire, College Lane, Hatfield AL10 9AB, UK
             }

   \date{}
   \authorrunning{Kundu, Vaidya, Mignone, Hardcastle}

 
  \abstract
   {Radio-loud AGNs are thought to possess various sites of particle acceleration, which gives rise to the observed non-thermal spectra.
    Stochastic turbulent acceleration (STA) and diffusive shock acceleration (DSA) are commonly cited as potential sources of high-energy particles in weakly magnetized environments.
    Together, these acceleration processes and various radiative losses determine the emission characteristics of these extra-galactic radio sources.}
   {The purpose of this research is to investigate the dynamical interplay between the STA and DSA in the radio lobes of FR-II radio galaxies, as well as the manner in which these acceleration mechanisms, along with a variety of radiative losses, collectively shape the emission features seen in these extra-galactic sources.}
   {A phenomenologically motivated model of STA is considered and subsequently employed on a magneto-hydrodynamically simulated radio lobe through a novel hybrid Eulerian-Lagrangian framework.}
   {STA gives rise to a curved particle spectrum that is morphologically different from the usual shock-accelerated spectrum.
    As a consequence of this structural difference in the underlying particle energy spectrum, various multi-wavelength features arise in the spectral energy distribution of the radio lobe. 
    Additionally, we observe enhanced diffuse X-ray emission from radio lobes for cases where STA is taken into account in addition to DSA.}
   {}


   \maketitle
%

\section{Introduction}
{Radio galaxies} are considered one of the most energetic systems in the universe.
These extra-galactic objects are observed to possess a huge reservoir of relativistic non-thermal particles, which collectively shape their emission features \citep{blandford_2019}.
Further, due to the abundance of highly energetic particles, these galaxies are generally considered a favourable site to study various high-energy phenomena \citep{meisenheimer_2003}. 
In recent years, with the advent of multi-messenger astronomy, different observations are uncovering various features and helping us understand different micro-physical processes happening in these systems \citep{marcowith_2020}. 

Low frequency radio observations of these radio galaxies provide insights about their morphological structures \citep[see][for more details]{hardcastle_2020}, their magnetic field strength \citep{Croston_2005b} and their age \citep{alexander_1987,carilli_1991,Mahatma_2019}.
Based on the brightness of these sources at $178$\,MHz, they are classified as Fanaroff Rilley (FR) class I (low-power) or II (high-power) \citep{fanaroff_1974}.
These two classes of radio galaxies are observed to manifest different morphological structures.
While FR-II sources exhibit a one-sided smooth spine-like structure with a bright termination point, FR-I sources show a two-sided plume-like structure.
Additionally, FR-II sources show prominent signs of turbulent cocoons that have an extent of a few hundred kiloparsecs and are often partly visible as lobes \citep[][]{mullin_2008,hardcastle_2020}. 
These lobes are believed to be highly magnetized cavities of rarefied plasma where most of the jet kinetic power is deposited.
Radio lobes also have a hotspot region near the jet termination region, responsible for accelerating particles to high energies via diffusive shock acceleration (DSA) \citep{brunetti_2001,prieto_2002,araudo_2018}.
These freshly shock accelerated particles further get mixed with the older plasma particles, already residing in the lobe, which makes the lobe a turbulent playground for various plasma waves to interact with the particles and eventually accelerate them via stochastic turbulent acceleration (STA). 
Such a mechanism has also been invoked to explain the particle acceleration in various astrophysical systems such as solar flares \citep{Petrosian_2012}, corona above accretion disk of compact object \citep{Dermer_1996,Liu_2004,Belmont_2008,Vurm_2009}, supernova remnant \citep{Bykov_1992,Kirk_1996,Macrowith_2010,Ferrand_2010},  gamma-ray burst \citep{Schlickeiser_2000}, emission from blazars \citep[see][and references therein]{Asano_2018,tavecchio_2022}, Fermi bubbles \citep{Mertsch_2019}, galaxy clusters \citep{brunetti_2007,donnert_2014,vazza_2021}.
STA has been invoked as a possible mechanism  for producing Ultra High Energy Cosmic Rays (UHECRs) from the radio lobe of Pictor A \citep{fan_2008} and Cen A \citep[][]{hardcastle_2009,sullivan_2009}.
Also, recently, it has been invoked as a plausible candidate in explaining the spectral curvature usually observed in FR-II radio lobes \citep[][]{harris_2019}.

In addition to the radio observations, X-ray observations of these radio loud AGNs have become popular due to the minimal contamination of the X-ray radiation by non-AGN sources.
Several components of these sources, such as radio lobes, hotspots, and collimated radio jet spine, are observed to radiate in the X-ray band \citep{vries_2018,Massaro_2018}.
Additionally, these lobes are often observed to give rise to diffuse X-ray emission from the region between the host galaxy and the radio hot spot, which is usually ascribed to the inverse Compton emission off the cosmic microwave background radiation (IC-CMB) \citep[][]{Hardcastle_2002,Croston_2005,blundell_2006}.
Recent observations reveal that the non-thermal X-ray emission from the radio lobe increases with red-shift, further supporting the IC-CMB origin \citep{gill_2021}.
Diffuse X-ray emission has also been reported in the jets of the FR-I class of radio galaxies and has been ascribed to a distributed particle acceleration mechanism \citep{hardcastle_2007b,worral_2008,worrall_2009}.
An IC-CMB model is also sometimes invoked to explain X-ray emission from the jets of FR-II radio galaxies and quasars, however such models require the jet to be highly relativistic and well aligned with the line of sight and consequently tend to imply very large physical jet lengths, sometimes in excess of several mega parsecs \citep{tavecchio_2000,celotti_2001,ghisellini_2005}. 
Further, recent polarimetric studies and high-energy gamma-ray constraints provide evidence supporting the synchrotron emission model as the origin of diffuse X-ray emission from AGN jets \citep[see][for a recent review]{perlman_2020}.
This consequently requires particles with very high energies to be present in the jet and also favours a distributed particle acceleration mechanism due to the short synchrotron lifetime of the radiating particles.

The present work explores, for the first time, the interplay of vital particle acceleration mechanisms in a weakly magnetised plasma environment such as the radio lobes of FR-II radio galaxies and studies their effect on the emission properties of these systems. 
Due to the complicated evolution of the dynamical quantities as a result of non-linear plasma flow pattern inside these lobes, we adopt a numerical approach for this work.
In particular, we have employed MHD simulations to produce radio lobes and analyse the emission features caused by particle energization in the presence of shocks and underlying turbulence.
We have adopted our recently developed second-order accurate STA framework \citep{Kundu_2021} for this purpose.
Owing to the increased computational complexity of the developed framework, this paper will focus on a 2D axisymmetric MHD jet model only while leaving the more computationally expensive 3D case to forthcoming works.

The paper is organised in the following way:
we describe our numerical setup for simulating a 2D axisymmetric AGN jet in section~\ref{sec:Dyn}.
Section~\ref{sec:emiss_setup} describes the numerical model to compute the emission properties.
In section~\ref{sec:results}, we present the results of the simulations.
In section~\ref{sec:summary} we summarise our findings and discuss the limitations of our model.

\section{Numerical Setup}\label{sec:setup}
In this section, we describe the numerical setup adopted for the present work. 
The radio lobes are typically associated with the termination point of the AGN jet, where the velocity of the jet material reduces considerably such that relativistic effects become negligible \citep{espinosa_2011}.
Further, as shown by \cite{hardcastle_2013}, numerical simulations of realistic radio lobe require high Mach number flows as well as very high-resolution meshes in order to have radio lobes in pressure equilibrium with the surrounding medium and to resolve the transverse radial equilibrium.
Therefore, to investigate the emission profile of the radio lobes, we focus on a non-relativistic scenario and perform a two dimensional axisymmetric ideal MHD simulation using PLUTO code \citep{mignone_2007}. In particular, we solve the following set of conservation equations,

\begin{equation}\label{eq:continuity}
  \frac{\partial\rho}{\partial t} + \nabla \cdot (\rho \vec v) = 0\,,
\end{equation}
\begin{equation}
  \frac{\partial\vec v}{\partial t} + (\vec v \cdot \nabla)\vec v = - \frac{1}{\rho}\nabla P 
    + \frac{1}{\rho}(\nabla \times \vec B)
 \times \vec B
     \,,
\end{equation}
\begin{equation}
  \frac{\partial P}{\partial t} + \vec v \cdot \nabla P
    + \Gamma P \nabla \cdot \vec v = 0 \,,
\end{equation}
\begin{equation}\label{eq:induction}
  \frac{\partial\vec B}{\partial t} = \nabla \times (\vec v \times \vec B)
     \,,
\end{equation}
The quantities $\rho$, $P$, $\vec v$ and $\vec B$ represents density, pressure, velocity and magnetic field respectively.
The magnetic field $\vec B$ further satisfies the constraint $\nabla \cdot \vec B=0$. 
$\Gamma$ represents the ratio of specific heats and its value is taken to be $5/3$, which is typically considered for supersonic non-relativistic jets \citep[][]{massaglia_2016}.
Eqs.~(\ref{eq:continuity})-(\ref{eq:induction}) is solved with HLLC Riemann solver using piece-wise linear reconstruction, van Leer flux limiter \citep{vanleer_1977} and second order Runge-Kutta time-stepping.
Additionally, we consider divergence cleaning \citep{dedner_2002} to satisfy the solenoidal constraint of magnetic field.

\subsection{Dynamical Setup} \label{sec:Dyn}
The two-dimensional axisymmetric simulations are carried out in a cylindrical geometry $\{r,z\}$ such that the radial and vertical extends range from $\{0,0\}$ to $\{65 L_0,195 L_0\}$ with a resolution of $780\times 2340$. 
The physical quantities defined in our simulations are appropriately scaled by defining length, velocity and density scales. 
For the length, we define the jet radius $r_{j} = L_{0} = 2$\,kpc as the scale length.
The core density is adopted as the scale for density such that $\rho_{0} = 5\times10^{-26}$\,gm/cc. 
Finally, for an ambient temperature $T_a = 2$\,keV, we define the sound speed $c_{a}=v_{0}=730$\,km/s as the scale velocity.

The ambient medium density is initialized with an isothermal King profile \citep[][]{king_1972},
\begin{dmath}\label{eq:king}
\rho_{a}=\frac{\rho_0}{\left(1+\left(\frac{R}{R_c}\right)^2\right)^{\frac{3\beta}{2}}},
\end{dmath}
where $\rho_{a}$ is the ambient density that consists of a core with radius $R_c=40 L_0$ and $R/L_0 = \sqrt{r^2+z^2}$ is the spherical radius. The value of power law index is kept constant as $\beta=0.35$.
Initially, the ambient medium is set in a hydrostatic equilibrium using a gravitational potential ($\Phi_{\rm k}$) \citep{krause_2005},
\begin{dmath}
\Phi_{\rm k}=\frac{3\beta k_{B}T_{a}}{2\mu m_{H}}\log\left(1+\left(\frac{R}{R_c}\right)^{\mathbf{2}}\right),
\end{dmath}
where $k_{B}$, $\mu$ and $m_{H}$ are the Boltzmann constant, mean molecular weight and mass of hydrogen atom respectively. 
The ambient pressure ($P_{a}$) is computed as follows,
\begin{dmath}\label{eq:prs}
P_{a} = \frac{\rho_{a}T_{a}k_{B}}{\mu}.
\end{dmath}
The ambient medium is set to be non-magnetized initially with the expectation that the magnetic field in the environment will have minimal impact on the non-thermal particle transport and the subsequent emission features within the lobe.

An under-dense beam of density $\rho_{j}=\eta\rho_{0}$ with velocity $v_{j}$ is continuously injected in the medium from a circular nozzle of radius $r_{j}$, along the vertical direction ($\hat{z}$) at $t=0$, with $\eta =0.1$ being the density contrast. 
The nozzle is placed within the numerical domain with a height of $0.5 L_0$. The adopted resolution samples the jet nozzle radius with 12 computational cells.
The injection velocity ($v_{j}$) is obtained by choosing the sonic Mach number $M$ such that,
\begin{equation}
    v_{j}=M c_{a},
\end{equation}
with $M = 25.0$. 
The injected beam includes a toroidal magnetic field ($B_{j}$) with the following radial profile \citep{lind_1989}
\begin{dmath}\label{eq:b_phi}
B_{ j,\phi} = \left\{\begin{array}{cl} 
B_{m}\frac{r}{r_{m}} & \mbox{$\text{for}\,\, r\leq r_{m}$}	\\
B_{m}\frac{r_{m}}{r} &\mbox{$\text{for}\,\, r_{m}\leq r\leq r_{j}$} \\
0 &\mbox{$\text{otherwise}$},
	\end{array} \right.
\end{dmath}
where the value of $B_{m}$ is governed by the plasma-beta parameter and $r_{m}$ is the magnetization radius. 
Such a magnetic field profile corresponds to a uniform current density within the radius $r_{m}$, zero current density between $r_{m}$ and $r$ and a return current at $r$. Further, such a configuration also respects the symmetry condition on the $z$ axis ($B_{j}=0$ at  $r=0$) \citep{komissarov_2007}. 
Additionally, a suitable gas pressure is provided inside the jet to ensure radial balance between the hoop stress and pressure gradient force,
\begin{dmath}\label{eq:prs_jet}
P_{j} = \left\{\begin{array}{cl} 
\left(\delta + \frac{2}{\kappa}\left(1-\frac{r^2}{r_m^2}\right)\right)P_{e} & \mbox{$\text{for}\,\, r< r_{m}$}	\\
\delta P_{e} &\mbox{$\text{for}\,\, r_{m}\leq r< r_{j}$} \\
P_{e} &\mbox{$\text{at}\,\, r=r_{j}$},
	\end{array} \right.
\end{dmath} 
where $\delta = 1-\frac{r_{m}^2}{\kappa r_{j}^2}$ and $\kappa = \frac{2P_e}{B_m^2}$ and $P_{e}$ is the pressure in units of $\rho_0 v_0^2$ at the nozzle radius, computed from the ambient medium ($P_{e}=P_{a}$ at $r=r_{j}$).
Owing to the constraint imposed by 2D axisymmetric geometry, the induction equation (Eq.~\ref{eq:induction}) does not enable conversion of the toroidal magnetic field ($B_{\phi}$) to a poloidal one.
As a result, we consider a minimal value of $B_m\sim 100$\,$\mu$G, to avoid significant amplification of the $B_{\phi}$ due to its continuous injection into the computational domain over time.
Further, the initial kinetic power of the jet is calculated from the quantities defined at the jet nozzle \citep{massaglia_2016},
\begin{dmath}
W = \frac{\pi}{2}\left(\frac{\Gamma k_{B}N_{A}}{\mu}\right)^{\frac{3}{2}}\eta \rho_{0} r_{j}^{2}M^{3}T_{a}^{\frac{3}{2}},
\end{dmath}
where $N_{A}$ is Avogadro's number. For the choices adopted in the present work, we obtain $W\simeq10^{45}$\,erg/s corresponding to the FR-II class of radio galaxies \citep{fanaroff_1974}.  

For the boundaries, we employ axisymmetric boundary condition about the axis for the inner $r$ boundary 
and free flow boundary conditions for all other boundaries in the computational domain. 

\subsection{Numerical setup to compute emission}\label{sec:emiss_setup}
The non-thermal emission from the radio lobe is modelled using the Eulerian-Lagrangian hybrid framework of the PLUTO code \citep{Vaidya_2018, mukherjee_2021}.
It employs passive Lagrangian (or macro-) particles whose dynamics is governed by the underlying fluid motion.
Physically, these macro-particles represent an ensemble of non-thermal particles (typically leptons) residing very closely together in physical space with a finite energy distribution.

The energy distribution of these macro-particles is evolved by solving the following transport equation,
\begin{equation}\label{eq:main}
  \DS    \pd{\chi_p}{\tau}
  + \pd{}{\gamma}\left[(S+D_{A})\chi_{p}\right] =
    \pd{}{\gamma}\left(D\pd{\chi_p}{\gamma}\right)
\end{equation}
where $\tau$ is the proper time, $\gamma\approx p/m_0 c$ is the Lorentz factor of the electrons, with $m_0$ being the rest mass of the electron and $c$ is the speed of light in vacuum. The dimensionless quantity $\chi_{p}=N/n$, with $N(p,\tau)$ being the number density of the non-thermal particles with momentum between $p$ and $p+dp$ and $n$ being the number density of the fluid at the position of the macro-particle. The quantity $S$ represents various radiative and adiabatic losses. The acceleration due to Fermi II order mechanism is given as $D_{A} = 2D/\gamma$ with $D$ being the momentum diffusion coefficient. We, for simplicity, neglect the source and sink terms in the transport equation.

Eq.~(\ref{eq:main}) is solved using a $2^{nd}$ order accurate finite-volume conservative implicit-explicit (IMEX) scheme \citep{Kundu_2021}.
The radiative losses considered include synchrotron, IC-CMB and adiabatic expansion to model the cooling processes of relativistic electrons. 
Additionally, as the particle spectra in the high energy region falls off rapidly due to various cooling processes, we follow \cite{winner_2019} and set the values of $\chi_{p} = 0$ beyond a threshold $\chi_{\rm cut}=10^{-21}$ .
Note that Eq.~(\ref{eq:main}) does not include shock acceleration; instead, a separate sub-grid prescription is employed to account for DSA \citep{Vaidya_2018,mukherjee_2021}.

The micro-physics of turbulent acceleration is encapsulated in the diffusion coefficient $D$. Typically, the empirical form of $D$ is given as an input in numerical simulations \citep{donnert_2014,vazza_2021}, as its quantification from first principles is complex particularly when applied to study large scale astrophysical environments. 
In this work we opt for a phenomenologically motivated ansatz of exponentially decaying hard-sphere turbulence as a model of STA inside the radio lobe.
We consider the acceleration timescale ($t_A$) as follows \citep{kundu_2022_conf}, 
\begin{equation}\label{eq:acc_time}
    t_{A} = \tau_{A} \exp\{(t-\tau_{t})/\tau_d\}
\end{equation}
where $\tau_{d}$ is the turbulence decay timescale, $\tau_{A}$ represents the acceleration timescale when turbulence decay is absent (or $\tau_{d}\to \infty$) and $t$ is the simulation time.
$\tau_{t}$ is the injection time of the macro-particle in a turbulent region. For a macro-particle that encounters a shock, its value is set to the time at which the last shock is encountered. While for those macro-particles that never undergo any shock, the value of $\tau_{t}$ is set to initial injection time in the computational domain.

This acceleration timescale has the capability to mimic the decay of turbulence, generally observed in various astrophysical sources.
The decay is a consequence of the finite lifetime of the turbulence and prevents particles from being continuously accelerated. 
For this work, we model $\tau_{A}$ and $\tau_{d}$ as follows,
\begin{align}
\label{eq:tau_a}
\begin{split}
  \tau_{A} &= \frac{\tau_{c}(\gamma_{\rm max} \rightarrow \gamma_{\rm min})}{\alpha},
\\
 \tau_{d} &= \tau_{A}.
\end{split}
\end{align}
where $\tau_{c}(\gamma_{\rm max} \rightarrow \gamma_{\rm min})$ represents the radiative loss time for a particle to cool from $\gamma_{\rm max}$ to $\gamma_{\rm min}$ and $\alpha$ is the ratio of synchrotron cooling time and acceleration time.
It also controls the efficiency of STA.
Larger $\alpha$ value corresponds to smaller $\tau_{A}$, STA timescale, and $\tau_{d}$, turbulence damping timescale.
Hence larger $\alpha$ indicates faster stochastic acceleration and faster damping.
Also, with lower values of $\alpha$, the effect of STA asymptotically diminishes.
It is a parametric representation that models the turbulence that actually occurs in realistic radio lobes of FR-II radio galaxies, which is unresolved in our simulation.
In this work, we vary its value and study how this affects the emission signatures.
The diffusion coefficient can subsequently be written as,
\begin{equation}
    D=\frac{\gamma^
    2\exp\{-(t-\tau_{t})/\tau_d\}}{\tau_{A}}.
\end{equation}
The $\gamma^{2}$ dependency of the diffusion coefficient is a characteristic of the hard-sphere turbulence.
Further, instead of such $\gamma^{2}$ dependent diffusion coefficient, one could also explore alternative diffusion models. 
For example, adopting Bohm type diffusion ($\propto \gamma$) could influence the results, however such a study of varying dependence of diffusion coefficient on $\gamma$ is beyond the scope of this paper.
To explore the ramifications of STA with varying efficiency on the emission of the simulated radio lobe structure, we use two alternative values for $\alpha=10^4$ and $10^5$ in this study.
Further, to sample the jet cocoon uniformly, we inject enough ($\sim 20$) macro-particles at every time step in the computational domain.
Initially, the normalized particle spectrum for each macro-particle is assumed to be a power-law, defined as $\chi_{p}(\gamma)=\chi_{0}\gamma^{-9}$, ranging from $\gamma_{\rm min}=1$ to $\gamma_{\rm max}=10^5$. 
The value of $\chi_{0}$ is set by prescribing the energy density of the macro-particles to be a fraction ($\approx 10^{-4}$) of the initial magnetic energy density.
Note that the  initial spectral index has a negligible effect on the emission of the system at later times as long as we consider a steep power-law. 

To compute the emissivity, we convolve the instantaneous energy spectrum of each macro-particle with the corresponding single particle radiative power and extrapolate it to the nearest grid cells. 
In particular we solve the following integral to compute the emissivity,
\begin{equation}\label{eq:emiss}
    j(\nu',n',\tau)=\int_{1}^{\infty}\mathcal{P}(\nu',\gamma',\psi')N'(\gamma',\tau)d\gamma' d\Omega',
\end{equation}
where, $\mathcal{P}(\nu',\gamma',\psi')$ is the power emitted by a non-thermal particle per unit frequency ($\nu'$) and unit solid angle ($\Omega'$) with lorentz factor $\gamma'$, and whose velocity makes an angle of $\psi'$ with the direction $n'$.
$N'(\gamma',\tau)$ is the number of micro-particles between lorentz factor $\gamma'$ and $\gamma'+d\gamma'$ at time $\tau'$.
In the case of an axisymmetric simulation, the magnetic field becomes independent of the polar angle and therefore to consider the line of sight (LOS) effect in the synchrotron emissivity, appropriate co-ordinate transformation is required \citep{meyer_2021}. We transform the magnetic field from cylindrical to Cartesian coordinates and compute the LOS effect by rotating the simulated structure explicitly.
The entire rotation (of $360^{\circ})$ is performed with an interval of $5^{\circ}$. 
Subsequently, the intensity maps of the structure are computed by doing a LOS integration of the calculated emissivity.
Note that all the emissivity calculation is performed by considering a viewing angle of $\theta = 90^{\circ}$ (i.e., along the $z=0$ plane in Cartesian co-ordinates).
\section{Results}\label{sec:results}

\begin{figure*}
    \centering
    \includegraphics[scale=0.4]{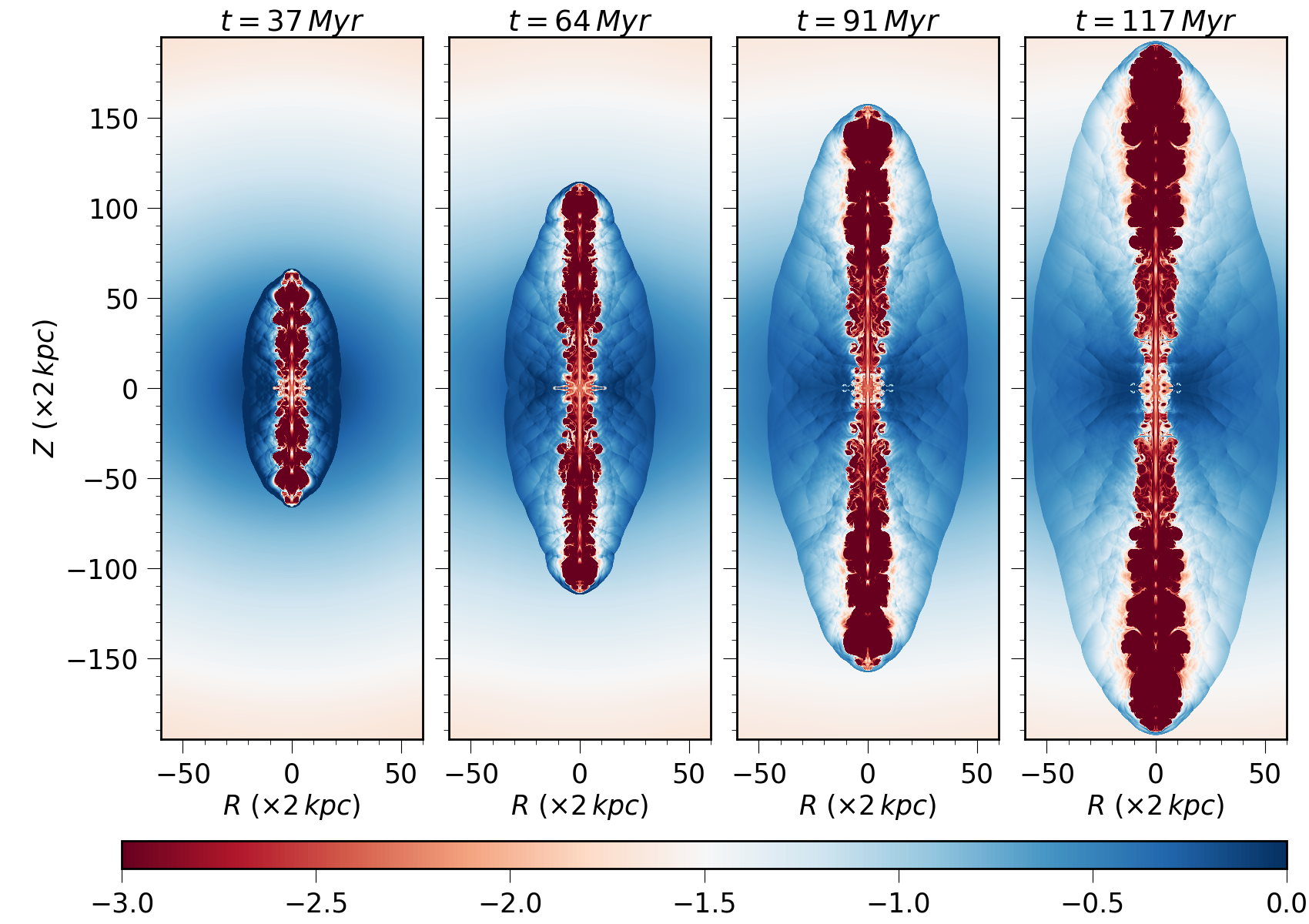}
    \caption{Normalized density $\rho/\rho_{0}$ evolution of the simulated radio lobe structure. The images depict a slice through the mid-plane of the notional 3D volume; all images are reflection-symmetric around the jet axis, and $z=0$ plane since the simulations are axisymmetric. The color-bar shows a logarithmic scale of density.}
    \label{fig:density}
\end{figure*}

\begin{figure*}
    \centering
    \includegraphics[scale=0.32]{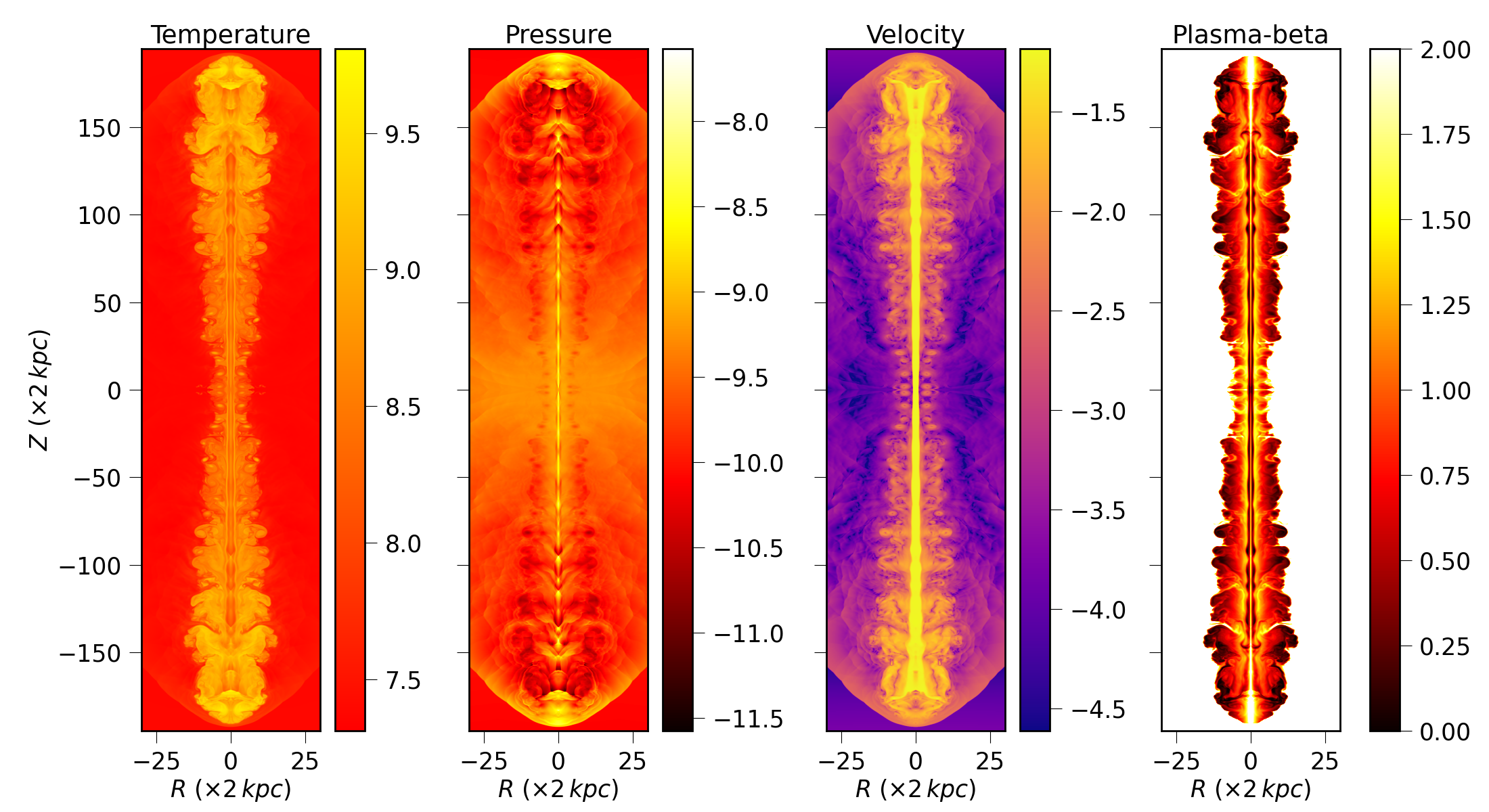}
    \caption{Temperature, pressure, absolute velocity and plasma-beta maps of the simulated jet structure for time $t=117$\,Myr. Temperature and pressure are shown in physical units; velocity is shown in the units of $c$ and the color-bars are shown in logarithmic scale. The average temperature of the radio lobe is of the order of $\sim70$\,keV, average plasma-beta is $\sim 32$ and average velocity is $\sim 0.02c$.}
    \label{fig:temp_prs}
\end{figure*}

\begin{figure*}
    \centering
    \includegraphics[scale=0.42]{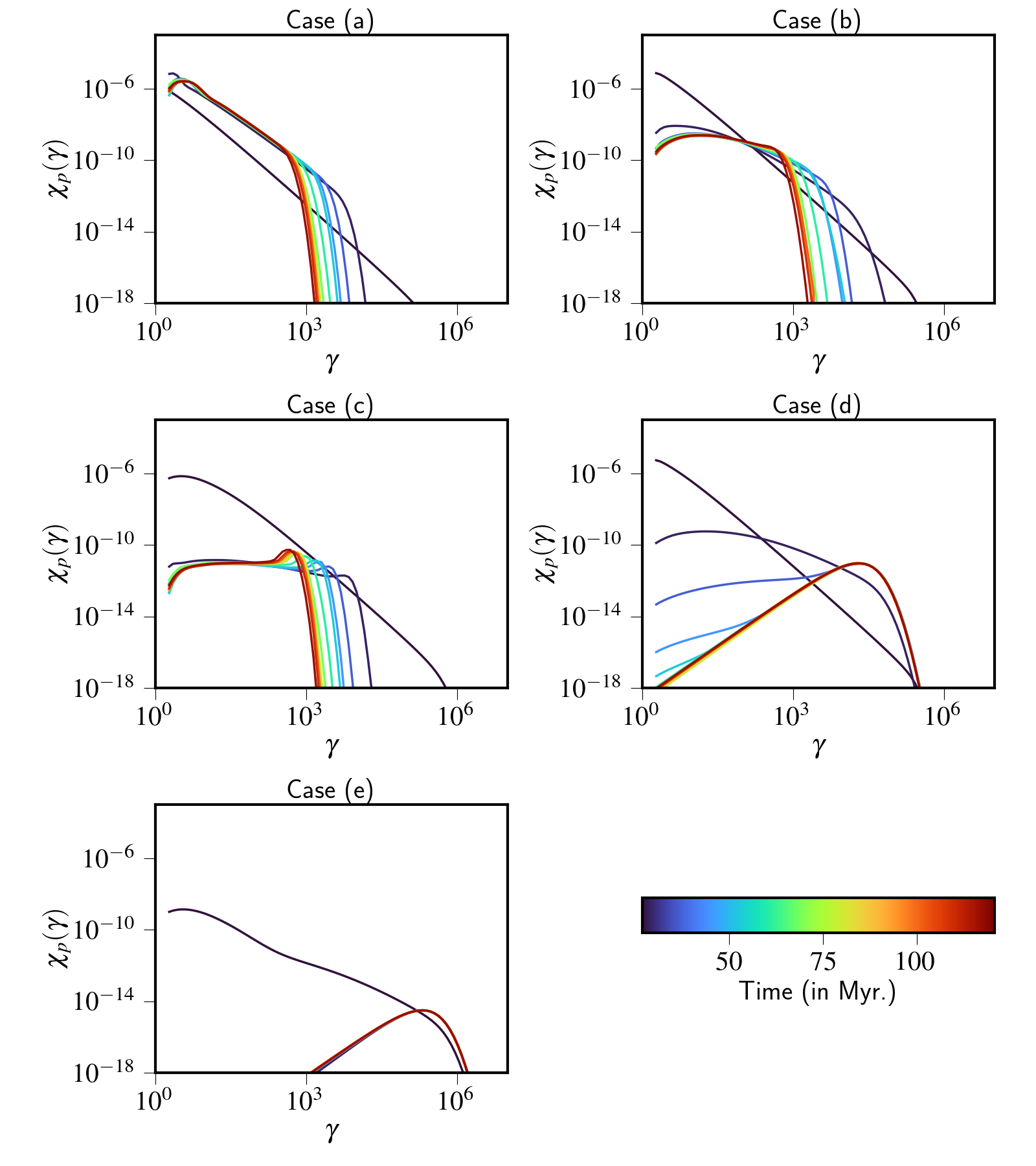}
    \caption{Evolution of the energy spectrum for a randomly chosen macro-particle for all the cases described in Table~\ref{tab:sim_cases}.
    The macro-particle encountered shock at a dynamical time of $t=25$\,Myr.
    The color-bar shows how much time has elapsed since the simulation began.
    The value of the lower end of the color-bar is set to the time when the macro-particle encountered the final shock.}
    \label{fig:spectrum}
\end{figure*}

\begin{figure*}
    \centering
    \includegraphics[scale=0.35]{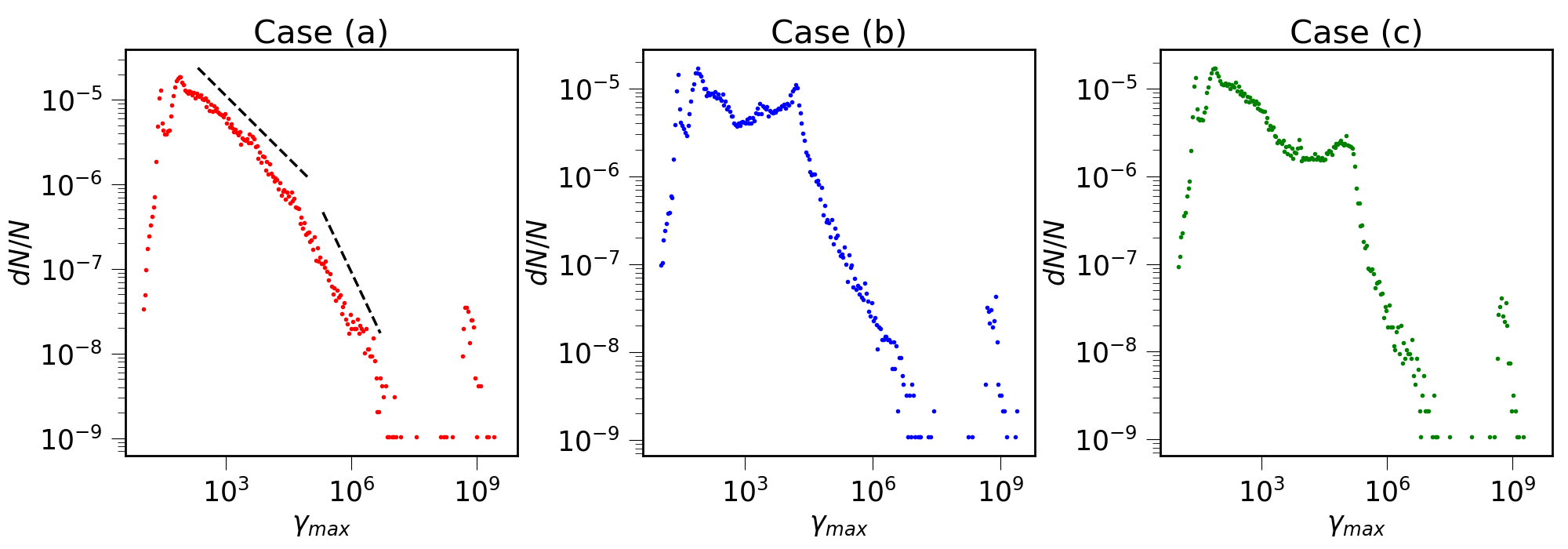}
    \caption{PDF of the cut-off energy for the entire macro-particle population. \textbf{Left, Middle } and \textbf{Right panel} shows the PDF for case (a), (b) and (c) respectively.}
    \label{fig:histogram}
\end{figure*}

We categorize the major results from our simulations in two parts. 
The first part gives an overview of dynamical aspects of radio lobes and the second part provides a detailed analysis of multi-wavelength emission signatures and particle acceleration processes within these lobes. 

\subsection{Dynamics}
%
We have carried out axisymmetric MHD simulations following the initial conditions described in Section~\ref{sec:setup} using the relevant jet and ambient medium parameters. The simulation is carried out up to a physical time of $\sim 120$\,Myr.
In Fig. \ref{fig:density}, we show the density evolution of the injected jets at different times, viz., t = $37$, $64$, $91$ and $117$\,Myr.
The density structure at every time snapshot shows an expanding bi-directional under-dense region which at a later time ($t = 117$\,Myr) can be identified as lobes \citep{english_2016}. 
Similar to \cite{hardcastle_2013}, we notice the formation of a long, thin lobe initially and a transverse expansion afterward.
This subsequent expansion in the transverse direction is attributed to the thermalization of the jet material by the shocks present in the lobe.
Further, we observe the formation of vortices at the lobe boundary, which are typically attributed to Kelvin-Helmholtz instabilities originated from the velocity shear between the lobe material and shocked ambient material.
Moreover, the entire structure is encapsulated within a forward-moving shock that can be seen to propagate through the ambient medium. This shock remains in the computational domain throughout the simulation time, preventing any mass, energy, and momentum from escaping the domain.

In Fig.~\ref{fig:temp_prs}, we show the temperature (left panel), thermal pressure (second panel), absolute velocity $|\vec{v}|$ (third panel) and plasma-beta (right panel) maps of the bi-directional jet at time $t = 117$\,Myr.
The temperature of the lobe (average value of $\sim70$\,keV) is relatively higher than the ambient medium ($T_a = 2$\,keV).
This is expected given the presence of a strong shock at the jet termination region, which is responsible for heating the jet material in the cocoon. 
The existence of the strong shock can further be seen from the pressure map as shown in the second panel of the figure.
The pressure map also provides evidence of multiple re-collimation shocks along the jet axis. 
Such shocks are expected to be favourable sites for accelerating particles via shock acceleration and are known to be a source of localised high-energy emissions.
Further, we observe the velocity of the jet is within the non-relativistic limit with an average value of $\sim 0.02c$.
The plasma-beta map, as depicted in the right panel of the figure, shows that the lobes are thermally dominated with an average lobe plasma-beta value of $\sim 32$.

The under-dense lobes observed in 2D simulations resemble the radio galaxies in a more consistent manner at later times \citep{hardcastle_2013}. 
In particular, when the expansion results in the length of the under-dense region being comparable to the core radius of the galaxy.

Therefore, in this work, for the emission studies, we adopt the dynamical results at time $t=117$\,Myr. 

\subsection{Emission}
%
We now turn our attention to the emission signatures of our model.
The discussion will be based on the comparison of synthetic emission signatures from different runs considered in our study. 
The parametric study focuses mainly on the properties of the stochastic turbulent acceleration mechanism. 
The details of these simulation runs are listed in Table~\ref{tab:sim_cases}, which considers various acceleration scenarios corresponding to different turbulent acceleration time scales $t_A$, while the background thermal fluid evolution remains exactly the same.

\begin{table*}
\centering
\begin{tabular}{| c  c  c  c  c  p{0.5\linewidth} |}
\hline\\
\textbf{Run ID} & \textbf{DSA} & \textbf{STA} & \textbf{Turbulent Decay} & \textbf{$\alpha$} & \textbf{{Remarks}} \\
\hline
\hline\\
Case (a) & YES & NO & NO & 0 & Energy spectrum exhibits power law with exponential cut-off; PDF of $\gamma_{\rm avg}$ shows power law; SED shows transient peaks.\\ \hline
Case (b) & YES & YES & YES & $10^4$ & Individual macro-particle energy spectrum exhibits curvature; $\gamma_{\rm max}$ PDF indicates accumulation of particles around $10^4$; $\gamma_{\rm avg}$ PDF exhibits low-energy cut-off. Peak radiation from synthetic SED is $10^{10}$\,Hz through synchrotron and $10^{19}$\,Hz via IC-CMB.\\ \hline
Case (c) & YES & YES & YES & $10^5$ & Individual macro-particle energy spectrum exhibits curvature. $\gamma_{\max}$ PDF shows particle accumulation around $10^{5}$; $\gamma_{\rm avg}$ PDF provides evidence of low-energy cut-off. Synthetic SED peak at $10^{13}$\,Hz through syncrotron and $10^{21}$\,Hz via IC-CMB.\\ \hline
Case (d) & YES & YES & NO & $10^4$ & Individual macro-particle energy spectrum exhibits steady ultra-relativistic Maxwellian structure peaking at $\gamma\approx10^4$.\\ \hline
Case (e) & YES & YES & NO & $10^5$ & Individual macro-particle energy spectrum exhibits steady ultra-relativistic Maxwellian structure peaking at $\gamma\approx10^5$.\\
\hline
\end{tabular}
\vskip2ex
\caption{Properties of the different cases, considered in the present study for calculating emission from the radio lobe. The first column contains labels for cases for further reference. The second, third, and fourth columns represent the presence or absence of DSA, STA, and turbulent decay effects on the emission runs. The fifth column depicts the value of free parameter $\alpha$ (Eq.\ref{eq:tau_a}) chosen for different runs and the last column describes the hey results for each of the cases.}
\label{tab:sim_cases}
\end{table*} 

The results obtained from cases (a) and (b) will be useful to comprehend the impact of STA and its interplay with DSA. 
Cases (b) and (c) will highlight the implications of having different turbulent decay timescales (see Eqs.~\ref{eq:acc_time}, ~\ref{eq:tau_a}). 
For cases (d) and (e), the turbulent decay is turned off by setting $\tau_{d}\rightarrow\infty$ in Eq.~(\ref{eq:acc_time}). 
Comparing results from these cases will demonstrate the effect of the turbulent decay process in our simulations. 
In realistic astrophysical environments, we expect the turbulence to decay on some time scale that is governed by the micro-physical properties of the wave-particle interaction in that system. 
As the current work incorporates turbulence via a sub-grid model, we have explored the implications of different parameters through these 5 cases.
All the results presented in this section are for a dynamical time of $117$\,Myr, unless specified otherwise.
Further, logarithmic binning has been adopted for all the histograms.

\subsubsection{Effect of Turbulent acceleration on individual macro-particle energy spectrum}
\label{sec:spectrum}
%
%
In Fig.~\ref{fig:spectrum}, we show the evolution of the energy spectra for all the cases listed in Table~\ref{tab:sim_cases} for a randomly chosen macro-particle which encountered final shock at a dynamical time $t=25$\,Myr.
In the simulations presented in this work, the majority of the Lagrangian macro-particles are observed to encounter more than one shock.
Among them, we selected this particular particle, which had experienced multiple shocks only at earlier times, as a representative candidate to demonstrate the effects of turbulent acceleration on the particle energy distribution in the downstream of the shock for all the case scenarios.
The effect of multiple shocks on the energy spectrum of a Lagrangian macro-particle without STA has already been investigated in the context of AGN jet simulation \citep[see][for example]{mukherjee_2021,giri_2022}.

The spectral evolution of the macro-particle of case (a) is shown in the top left panel.
The spectrum exhibits a power-law with a high-energy cut-off which gradually shifts to lower energy with time, owing to various energy losses.
Additionally, a small hump can be seen in the low-energy part of the spectrum which is due to an excess of lower energy electrons arising from their higher energy counterparts due to radiative cooling.

The shape of the spectrum changes considerably when STA is considered in addition to DSA.
For cases (d) and (e) (shown in the right plot of the middle panel and left plot of the bottom panel respectively), the spectrum exhibits an ultra-relativistic Maxwellian distribution at later times.
This is a consequence of a steady competition between stochastic acceleration and radiative losses resulting in the acceleration of low-energy electrons towards higher energies \citep{Kundu_2021}.
Moreover, the peak of the distribution corresponds to the value of $\gamma$ at which acceleration and loss time scales match, i.e.,  $\tau_{c}=t_{A}$. 
We find that the peak corresponds to $\gamma \approx \alpha$ and it depends on the choice of the turbulent acceleration timescale (see Eq.~\ref{eq:acc_time}).

When turbulent decay is included (cases b and c) we observe flatness of the spectrum in the lower energy regime, as compared to the power-law behavior observed in case (a), along with a high energy cut-off.
The flattening of the lower energy component of the spectrum is a consequence of the fact that STA provides a continuous acceleration to all the micro-particles, resulting in their acceleration to higher energies, depopulating the low energy regime. 

Further, we point out that for the macro-particles which have encountered a shock, STA starts acting in the downstream and it modifies the energy spectra on a time-scale that depends on $t_A$ (Eq.~\ref{eq:acc_time}) which in turns is regulated by the turbulent decay time-scale $\tau_d$ and consequently develops a cut-off that moves towards lower energies.

In summary, the spectral evolution of a macro-particle, presented in Fig.~\ref{fig:spectrum} for different cases, clearly indicate that the presence of turbulent acceleration significantly affects the spectral energy distribution and its evolution.
Our results indicate that, in the absence of turbulent decay, spectral evolution eventually relaxes towards a steady-state configuration in which energy losses are balanced by turbulent acceleration, while, on accounting for the decay of turbulence, the energy spectrum exhibits a non-stationary behaviour in time and the cut-off is governed by the radiative loss time scale subsequent to the decay of turbulence.
Further, the spectrum shows flattening in the lower energy regime owing to the energization of low energy micro-particles to higher energy by STA.

\subsubsection{Effect of Turbulent acceleration on particle population} 
\label{sec: collective}
%
%
This section focuses on the effects of turbulent acceleration on the entire macro-particle population in the lobe.
In particular, we compute the effect of the STA with turbulence decay on the cut-off energy ($\gamma_{\rm max}$) for the macro-particle population.
To compute the cut-off energy of a macro-particle we consider a generic form of its energy spectrum,
\begin{equation}\label{eq:fit}
    \gamma^{-m}\exp\left(-\frac{\gamma}{\gamma_{\rm max}}\right),
\end{equation}
where $m$ can be positive or negative depending on the macro-particle and $\gamma_{\max}$ is the cut-off energy.
The exponential decay term takes care of the effects on the spectrum due to various radiative losses (see section~{\ref{sec:spectrum}}).
The value of $\gamma_{\max}$ is calculated by multiplying Eq.~(\ref{eq:fit}) by a power-law profile, $\gamma^{10}$, and calculating the maximum point of the resultant curve.

In Fig.~\ref{fig:histogram}, we show the probability distribution function (PDF) of the maximum (or cut-off) energy ($\gamma_{\max}$) attained by individual macro-particles for cases (a) (left panel), (b) (middle panel), and (c) (right panel). 
For case (a), the distribution peaks around $\gamma_{\max}\approx 10^2$, followed by a broken power-law like tail beyond that. 
The origin of this peak can be attributed to the presence of various radiative losses in the system.  
The peak is also observed to gradually move towards lower values of $\gamma_{\max}$ with time.
To support the above argument, we undertake the following exercise:
for a particle undergoing synchrotron cooling only, the initial Lorentz factor $\gamma^{'}$ after a time period of $t^{'}$ becomes,
\begin{equation}\label{eq:fin_gam}
    \gamma^{*}=\frac{1}{C_{0}B^2 t^{'}+\frac{1}{\gamma^{'}}}
\end{equation}
with $C_{0}=1.28\times 10^{-9}$ is the synchrotron constant for electron, $B$ is the magnetic field.
For our case, considering an averaged magnetic field of $B=19.70\,\mu$G and $t^{'}=117$\,Myr we get $\gamma^{*}\approx 5.4\times 10^{2}$ for a range of $\gamma^{'}$ values, which correlates with the position of the peak.
The break in the power-law around $\gamma_{\max}\sim 10^{5}$ is attributed to the continuous injection of the macro-particles in the computational domain with $\gamma_{\max}=10^{5}$ (see section \ref{sec:emiss_setup}).
The presence of an additional smaller peak around $\gamma_{\max}\sim10^{9}$ can also be observed.
This smaller peak is a transient feature, which arises from recently shocked macro-particles and is a manifestation of the continuous injection of jet material along with the Lagrangian macro-particles inside the computational domain.
The presence of this transient peak has been reported in earlier works as well \citep[see for example][]{borse_2021}. 
Further, the power-law trend of the tail of the PDF is typically ascribed to the interplay between the continuous injection of macro-particles in the computational domain and the shock acceleration of these freshly injected particles. Such a power-law like behaviour of the distribution in an AGN jet cocoon has also been reported in \cite{mukherjee_2021}.

The PDFs for cases (b) and (c) show some additional peaks, as compared to case (a). 
The origin of the peak at $\gamma_{\max}\sim 10^{2}$ is similar to case (a), while the high-energy one
($\gamma_{\max}\sim 10^{9}$) is again due to recently shocked macro-particles. 
In addition, one can observe humps at $\gamma_{\max} \sim 10^{4}$ (for case b) and at $\gamma_{\max} \sim 10^{5}$ (for case c).
Their presence is caused by particles undergoing turbulent acceleration downstream of the shock, resulting in freezing the evolution of the cut-off at $\gamma_{\max}\approx\alpha$ for some time due to the competition between STA and radiative losses and afterwards, due to the decay of turbulence, the cut-off continues to decrease towards lower energy as dictated by loss processes.

\begin{figure*}
    \centering
    \includegraphics[scale=0.35]{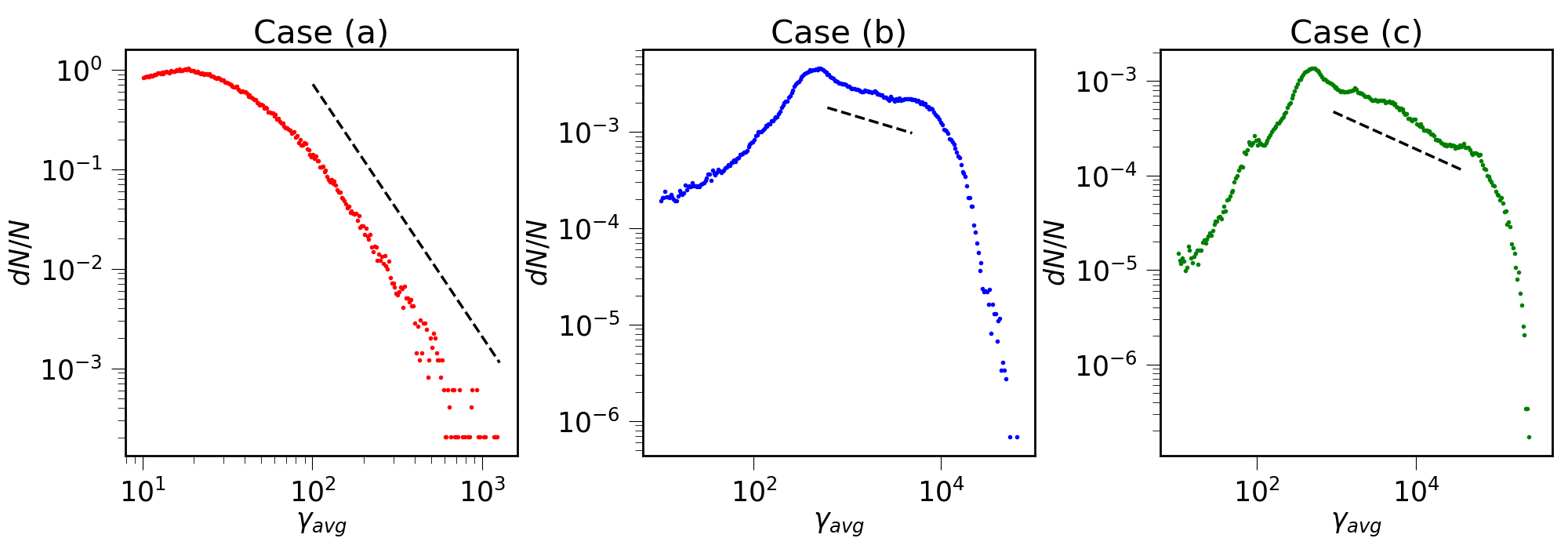}
    \caption{PDF of the $\gamma_{\rm avg}$ (see Eq. \ref{eq:gamma_avg}) for the entire macro-particle population. \textbf{Left, Middle } and \textbf{Right panel} shows the PDF for case (a), (b) and (c) respectively.}
    \label{fig:aver_gamma}
\end{figure*}
\begin{figure}[!h]
    \centering
    \includegraphics[scale=0.35]{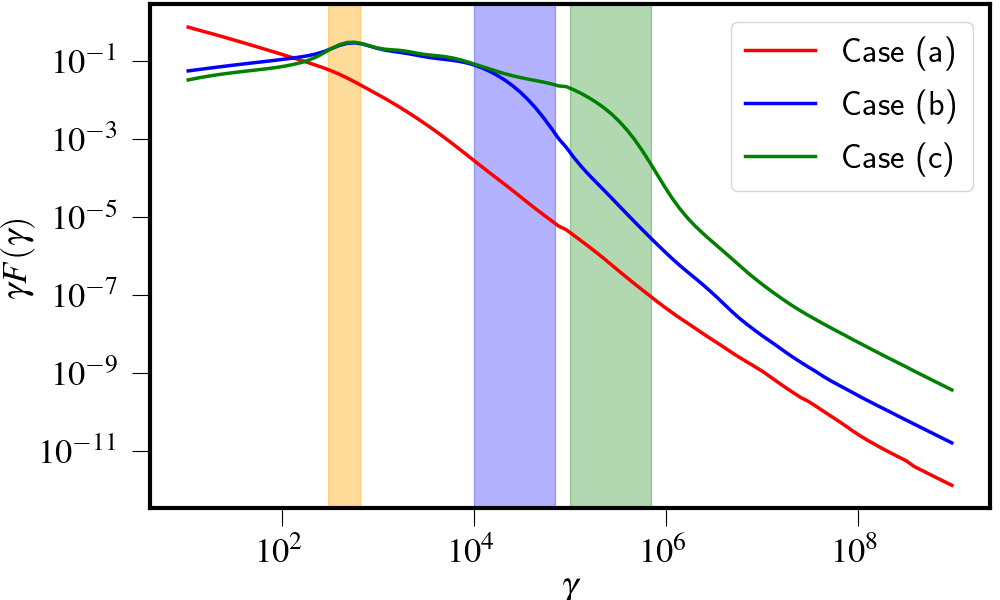}
    \caption{Integrated spectrum of the entire macro-particle population for all the three cases.
    The portion of the spectrum highlighted in orange corresponds to the low energy break.
    The highlighted portion of the spectrum by blue (and green) corresponds to the high energy cut-off for case (b) (and case (c)).}
    \label{fig:integrated}
\end{figure}
To understand the distribution of electron energy within macro-particles, we also estimate the average value of  $\gamma$ (at the final simulation time, $t = 117$ Myr) denoted by $\gamma_{\rm avg}$ as: 
\begin{equation}\label{eq:gamma_avg}
    \gamma_{\rm avg}(t) = \frac{\int_{\gamma_{min}}^{\gamma_{\rm max}} \gamma N(\gamma,t)d\gamma}{\int_{\gamma_{min}}^{\gamma_{\rm max}} N(\gamma,t)d\gamma}\,.
\end{equation}
where $\gamma_\max$ and $\gamma_\min$ are given in section \ref{sec:emiss_setup}.
In Fig.~\ref{fig:aver_gamma}, we plot the PDF of $\gamma_{\rm avg}$ for the entire macro-particle population. 
In the left panel of the figure, we show the PDF for $\gamma_{\rm avg}$ for case (a).
The distribution exhibits a power-law tail ($\propto \gamma_{\rm avg}^{-q}$, with $q\approx2.54$) beyond $\gamma_{\rm avg}\sim 10^{2}$.
For cases (b) and (c) (depicted in the middle and right panel of the figure), the PDFs exhibit a power-law distribution starting from $\gamma_{\rm avg}\sim 10^{3}$ with a small hump and an exponential cut-off.
The hump feature arises due to competition between STA and radiative losses (see above).
It is interesting to note that the slope of the power-law for both the cases (b) and (c) ($q=0.29$, $0.38$ respectively) is flatter than case (a).
This is a consequence of the fact that STA continuously supplies energy to the macro-particles by accelerating the low energetic micro-particles to the higher energy, thus compensating for the radiative losses, as opposed to the case with only DSA.
Finally, in presence of both DSA and STA, the $\gamma_{\rm avg}$ PDFs exhibit a low-energy break around $\gamma_{\rm avg}\sim10^{3}$ owing to the fact that STA boosts low-energy particles to higher energies.
This process is absent if only DSA is present, since there is no selective mechanism to accelerate only the low-energy particles during shock acceleration (which involve convolution of the entire upstream spectrum of each macro-particles to downstream \citep{mukherjee_2021}) and hence $\gamma_{\rm avg}$ PDF cannot form a low-energy break.

Further, in Fig.~\ref{fig:integrated}, we present the integrated particle spectrum considering the whole macro-particle population for each of the three case scenarios.
The integrated particle spectrum is calculated as follows:
\begin{equation}
    F(\gamma)=\sum_{i}\frac{\chi^{i}_{p}(\gamma)}{\mathcal{N}_{i}(\gamma)\int\chi^{i}_{p}(\gamma')d\gamma'}\,,
\end{equation}
where $i$ corresponds to individual macro-particles inside the computational domain, $\chi^{i}_{p}(\gamma)$ is the distribution function of the $i^{th}$- macro-particle and $\mathcal{N}_{i}(\gamma)$ represents the number of macro-particles with Lorentz factor $\gamma$.
The DSA spectrum (case (a)) is in the form of a broken power-law with the break at $\gamma \approx 5 \times 10^2$ (region highlighted with orange color in the figure).
Such a behaviour is expected when computing a resultant distribution comprising all the macro-particles, where the spectral evolution is mediated by shock acceleration and radiative losses \citep{heavens_1987}.
The position of the break has a direct correspondence with the peak in the $\gamma_{\max}$ PDF for case (a) and can be explained by the same reasoning (see Eq.~\ref{eq:fin_gam}).
When STA is taken into account (cases (b) and (c)), the spectrum exhibits an inverse power-law behaviour for $\gamma \lesssim 4\times 10^2$, followed by a low energy break and a power-law trend with a high energy cut-off highlighted in blue and green for cases (b) and (c), respectively in the figure.
The spectral behaviour in the region $\gamma\lesssim 4\times10^2$ is a manifestation of the low-energy flattening in the individual macro-particle spectrum (see section~\ref{sec:spectrum}) due to turbulent acceleration.
The origin of the low energy break bears a similar explanation as the case (a). However, for cases where STA is taken into account, the cut-off is accompanied by piled up micro-particles (see case (c) of Fig.~\ref{fig:spectrum}) as opposed to case (a), which is why the break appears more prominent in cases (b) and (c).
The high energy cut-off in the integrated particle spectrum (at $\gamma\approx10^4$ for case (b) and $\approx10^5$ for case (c)) is governed by the formation of the quasi-stationary cut-off in the individual macro-particle spectrum due to the interplay of DSA and STA.
As a result, the position of these high energy cut-offs has an exact correspondence with the peaks observed in Fig.~\ref{fig:histogram} for the cases where STA is taken into account.
The power-law trend beyond $\gamma\gtrsim 10^6$ for all the case scenarios is a consequence of the continuous macro-particle injection in the computational domain and a fraction of them subsequently getting shock accelerated.

In summary, turbulent acceleration with exponential decay modifies the macro-particles' maximum energy ($\gamma_{\rm max}$) distribution by presenting additional hump to the PDFs.
The location of the humps is closely connected to the $\gamma$ of individual macro-particles where $\tau_{c}=t_{A}$.
The PDF of $\gamma_{\rm avg}$ for cases (b) and (c) exhibits a power-law trend with an exponential cut-off and a low energy break.
The integrated spectrum with only DSA exhibits a low energy break, whereas with STA, an additional cut-off at high energy is also seen.

\begin{figure*}
    \centering
    \includegraphics[scale=0.4]{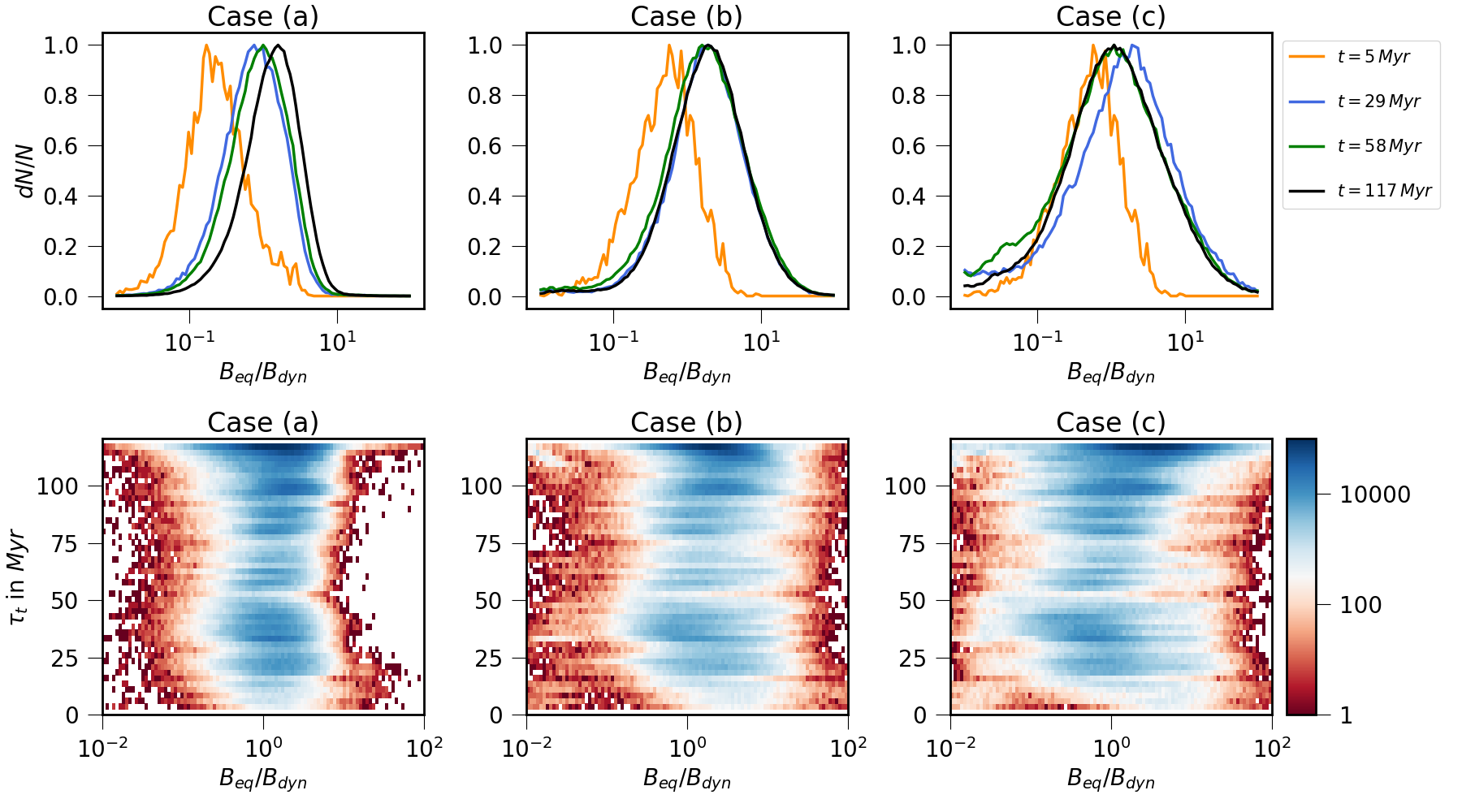}
    \caption{Histogram of macro-particles with respect to $B_{\rm eq}/B_{\rm dyn}$ to further study effect of STA on the macro-particle population.
    Histograms are normalized and then scaled with the maximum value.
    \textbf{Top panel:} Shows the histograms for 3 different cases at 4 different times (shown in different colors). \textbf{Top left, middle } and \textbf{right panel} shows the histogram for case (a), (b) and (c) respectively. \textbf{Bottom panel:} 2D histograms showing $\tau_{t}$ vs. $B_{\rm eq}/B_{\rm dyn}$ at the final time $t=117$\,Myr for 3 cases. \textbf{Bottom left, middle } and \textbf{right panel} shows the histogram for case (a), (b) and (c) respectively. The colorbar at the bottom panel shows the number of macro-particles. }
    \label{fig:equipartition}
\end{figure*}

\subsubsection{Turbulent acceleration as a sustained acceleration process}
%

In this section we examine how STA supports the macro-particles to sustain their energy from extreme radiative losses.
To properly characterize this behaviour we consider an equivalent magnetic field for each macro-particle and compare it with the dynamical magnetic field at the position of the macro-particle.
This is computed from the instantaneous single macro-particle energy distribution as follows,
\begin{equation}\label{eq:mag_equip}
    \frac{B_{\rm eq}^{2}}{8\pi}=m_{0}c^2\int_{\gamma_{min}}^{\gamma_{\rm max}}\gamma N(\gamma,t)d\gamma,
\end{equation}
where $B_{\rm eq}$ is the corresponding equivalent magnetic field. 

Following Eq.~(\ref{eq:mag_equip}), we compute $B_{\rm eq}$ for cases (a), (b), and (c) and compare it with the corresponding dynamical magnetic field computed at the local macro-particle position at each instant, $B_{\rm dyn}$.
We plot the time evolution of the histogram of the quantity, $B_{\rm eq}/B_{\rm dyn}$, on a logarithmic scale, for all three cases in the top panel of Fig.~\ref{fig:equipartition}, where orange, blue, green, and black curves in each panel depict the histogram at times $5$\,Myr, $29$\,Myr, $58$\, Myr, and $117$\, Myr, respectively.
All the histograms are normalized so that the maximum peak value is unity.

As shown in the top left panel, for case (a), the histogram gradually shifts towards a state with, $B_{\rm dyn} \sim B_{\rm eq}$ as time progresses.
The other two cases (b) and (c) exhibit a similar pattern, where one observes a broadening of the histogram as well.
For case (a), the shape of the PDF
can be observed to 
evolves to a negatively skewed distribution on a logarithmic scale. 
To analyze the reason behind this kind of evolution, we show (bottom left panel) a 2D histogram depicting the value of $\tau_{t}$ with respect to the magnetic field ratio, which indicates implying  that the macro-particle population with a larger magnetic field ratio has recently been shocked.
This should not be surprising since the shock acceleration energizes particles thereby increasing $B_{\rm eq}$. 
The 2D histogram also shows that a relatively small
fraction of macro-particles has magnetic field ratios larger than unity, due to the absence of any further acceleration process.
As a result these particles undergo strong cooling and quickly lose their energy, hence featuring an exponential fall in the histogram beyond $B_{\rm eq}\sim B_{\rm dyn}$ (top left plot of Fig.~\ref{fig:equipartition}). 

On the contrary, for cases (b) and (c) (top middle and right panels), the 1D histogram evolves to a more extended structure, which closely resembles the log-normal shape.
This extended form of the histograms is ascribed to the presence of STA which provides a continuous acceleration to the macro-particles and helps them maintain their energy even in the presence of radiative cooling. 
This is further confirmed by observing the corresponding 2D histograms in the bottom panels (middle and right, respectively).
In contrast to case (a), both figures show more macro-particles in the region $B_{\rm eq}/B_{\rm dyn}\gtrsim 1$. 
Also we can infer that even macro-particles that were shocked  earlier (smaller $\tau_{t}$) feature a higher value of  $B_{\rm eq}/B_{\rm dyn}$ as a result of the fact that with STA macro-particles can sustain their energy for a longer amount of time.

In summary, for all the cases, one observes the distribution gradually evolve towards a state where $B_{\rm eq}\sim B_{\rm dyn}$. 
Further, due to the presence of STA, as compared to only DSA, the histogram manifests a more extended structure that is evenly spread due to the macro-particles which were shocked at earlier time but could sustain their energy from radiative losses because of STA. 

\begin{figure}
    \centering
    \includegraphics[scale=0.26]{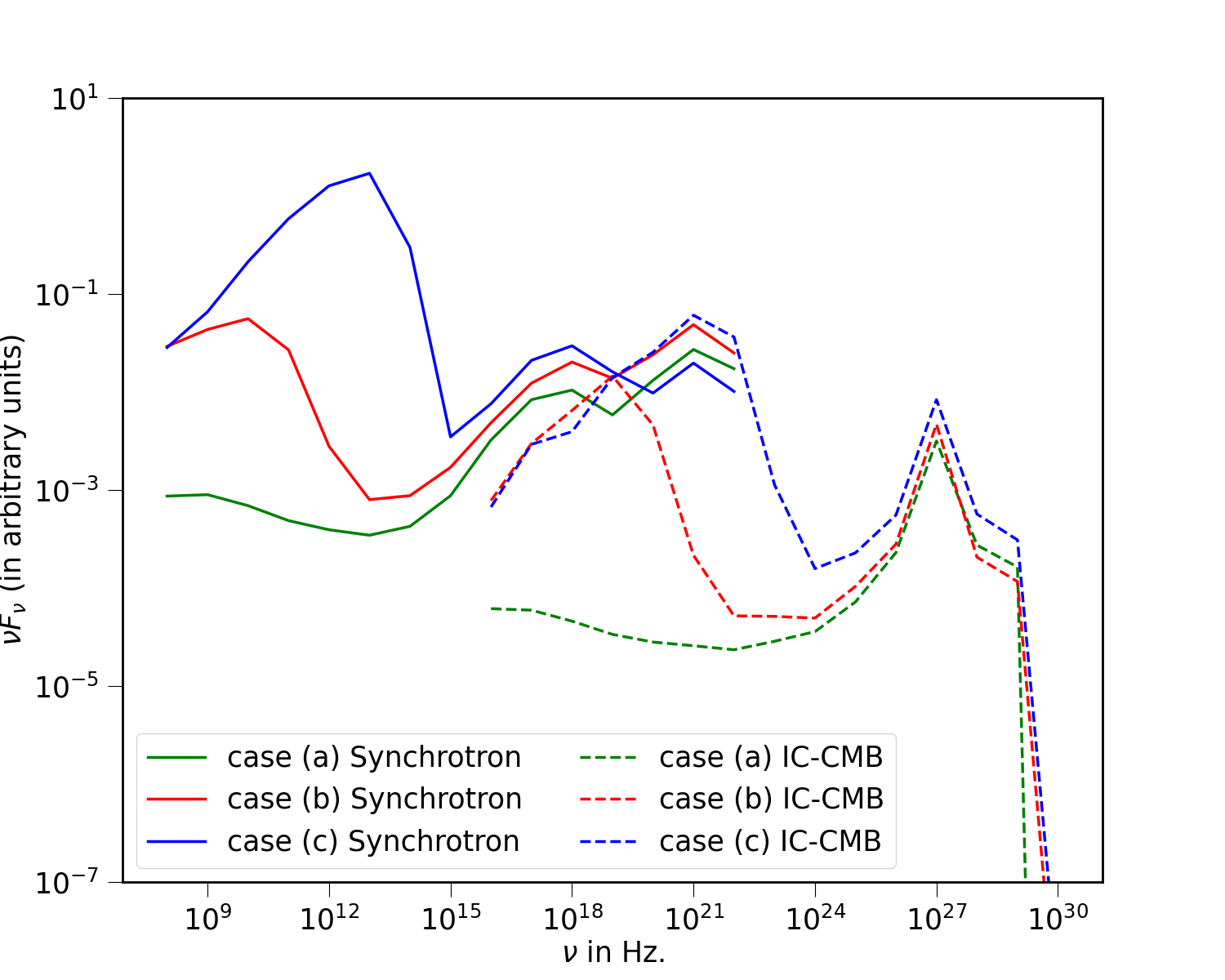}
    \caption{Synthetic spectral energy distribution (SED) for case (a) (shown in green), case (b) (shown in red) and case (c) (shown in blue). The SED due to synchrotron mechanism is shown in solid lines and the IC-CMB part is shown in dashed lines. 
    The vertical axis shows the value of $\nu F_{\nu}$ in arbitrary units.}
    \label{fig:sed}
\end{figure}

\begin{figure*}
    \centering
    \includegraphics[scale=0.66]{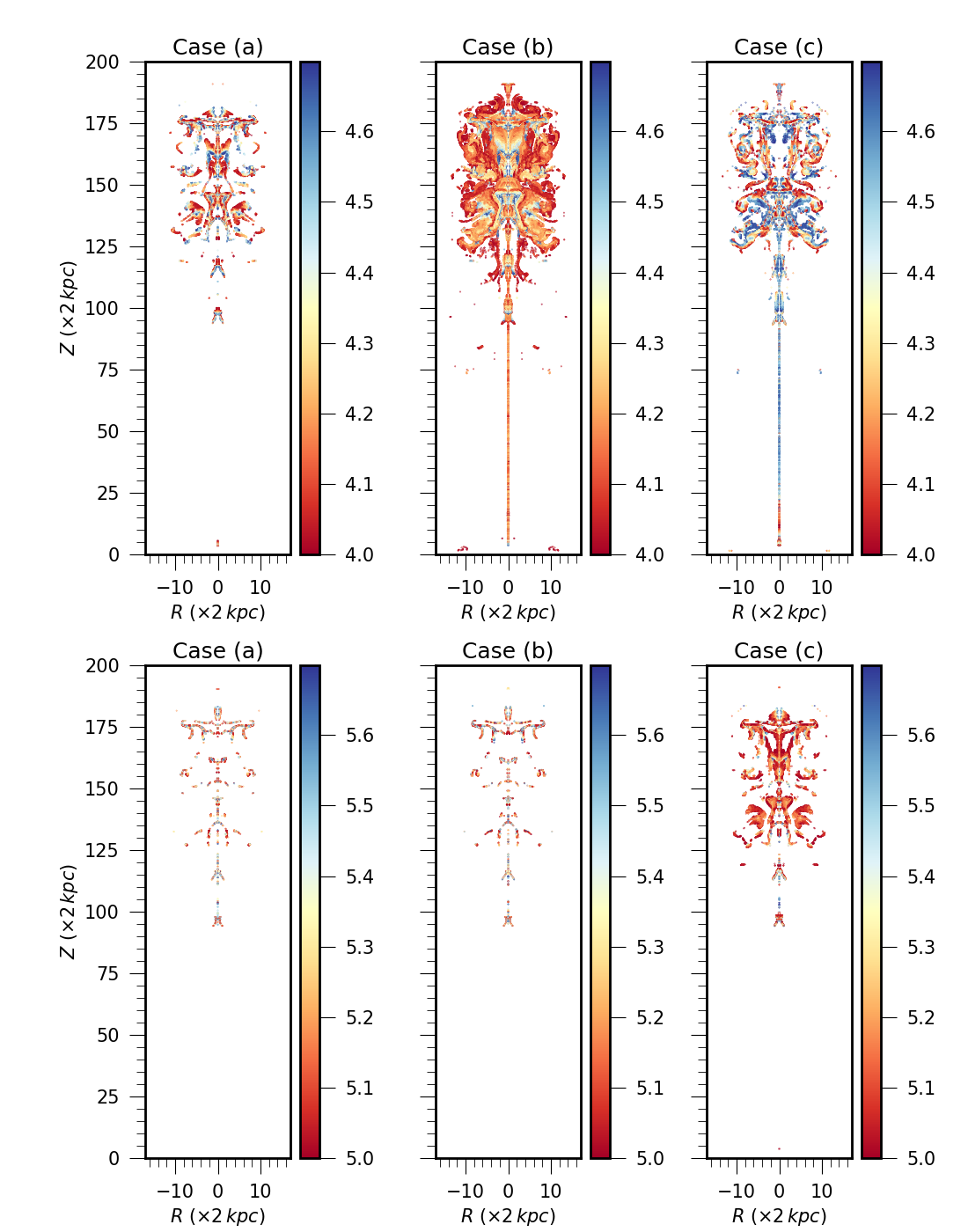}
    \caption{Spatial distribution of the particles responsible for the peaks in SED. \textbf{Top panel:} shows the position of the particle population with $\gamma_{\rm max}\sim 10^4$ for case (a) (\textbf{left}), case (b) (\textbf{middle}) and case (c) (\textbf{right}); \textbf{Bottom panel:} 
    shows the position of the particle population with $\gamma_{\rm max}\sim 10^5$ for case (a) (\textbf{left}), case (b) (\textbf{middle}) and case (c) (\textbf{right}).}
    \label{fig:gam_spatial}
\end{figure*}

\subsubsection{Synthetic Spectral Energy Distribution of Radio lobe}
%
In Fig.~\ref{fig:sed}, we present the spectral energy distribution (SED) for cases (a) (green line), (b) (red line) and (c) (blue line).
The SED is calculated by integrating the emissivity (Eq.~\ref{eq:emiss}) along the line of sight \citep{Vaidya_2018} with two different radiation mechanisms: synchrotron (solid lines) and IC-CMB (dashed lines).
The synchrotron SED shows, for case (a), enhanced emission in the X-ray band with multiple peaks at  $\nu \sim 10^{18}$  and $\nu \sim 10^{21}$  Hz. 
These peaks are originated from freshly shocked macro-particles \citep{borse_2021, mukherjee_2021}. 
This can be further verified analytically using the relation between the critical (or cut-off) frequency ($\nu_{c}$) of synchrotron radiation and the corresponding $\gamma$ \citep[see for example Eqs.~(5.80) from][]{condon_2016},
\begin{equation}
    \nu_{c}\approx\frac{\gamma^2 eB}{2\pi m_{e}c},
\end{equation}
For instance, with an averaged magnetic field of the lobe $B= 19.70$\,$\mu$G and $\nu_{c} \sim 10^{21}$\,Hz, one obtains a corresponding value for $\gamma \sim 10^9$, which is consistent with the peak in the PDF of $\gamma_{\max}$ as seen in Fig.~\ref{fig:histogram}. 

For case (b), in addition to similar shock-induced transient signatures, the synchrotron emission shows a distinct peak in the low energy GHz radio band ($\nu\sim 10^{10}$ Hz).
The origin of such a low energy peak is a direct evidence of turbulent acceleration and corresponds to the hump in the PDF at $\gamma_{\max} \sim 10^4$ (see middle panel of Fig.~\ref{fig:histogram}).
Likewise, the synchrotron peak can also be observed for case (c) at a slightly higher energy, $\nu\sim10^{13}$\,Hz. 
The macro-particles that are accelerated via STA and give rise to the peak in PDF around $\gamma_{\rm max}\sim 10^{5}$ (right panel of Fig.~\ref{fig:histogram}) are mainly contributing to the emission at this frequency band.
The macro-particle population that is stochastically accelerated in cases (b) and (c) is not only responsible for synchroton emission but also contributes to the distinct peaks in the IC-CMB spectral energy distribution ($\nu\sim 10^{19}$\,Hz for case b, $\nu\sim 10^{21}$\,Hz for case c).
We have verified that these correspond to the frequency of the photons scattered of a population of electrons with energy $\gamma_{\max}\sim10^{4}$ and $\gamma_{\max}\sim10^{5}$ for case (b) and (c), respectively.
In fact, the post-scattering frequency of the photons $\nu_s$ is related to the electron energy as follows:
\begin{equation}\label{eq:iccmb}
    \nu_{s}\approx \gamma_{\max}^{2}\nu_{0}\,,
\end{equation}
where $\nu_{0}$ is the frequency at which the cosmic microwave background (CMB) radiates. 
Using $\nu_{0}=160$\,GHz in Eq. ({\ref{eq:iccmb}}), we find that an electron population at $\gamma_{\max}\sim 10^4$ would scatter the CMB photons at a frequency of $\sim10^{19}$\,Hz. 
A similar inference can be drawn for the origin of the IC-CMB peak around $\sim 10^{21}$\,Hz for case (c). 
Additionally, the peaks in the $\gamma$-ray band ($\nu \sim 10^{27}$\,Hz) for all cases corresponds to the particles with $\gamma_{\max}\sim 10^9$.

After observing the SED and identifying the particle populations responsible for the various peaks, we proceed to show the spatial distributions of these particle populations in order to understand the resulting emission structure.
In Fig.~\ref{fig:gam_spatial} we show the spatial distribution of the particle populations responsible for the aforementioned peaks.
The top panels depict the particle distributions with $\gamma_{\max}\sim10^4$ for case (a) (left plot), case (b) (middle plot) and case (c) (right plot).
These particles are correlated to the peak in the SED caused by IC-CMB at $\nu\sim10^{19}$\,Hz, as explained earlier in this section.
The macro-particles in case (a) can be seen to be more confined around the shocks in the beam and, to a lesser extent, to the cocoon region.
The reason for this is that, after the shock acceleration, the macro-particles energy evolution is governed by loss mechanisms only and as a result, they lose a considerable amount of energy in a short distance.
On the contrary, when turbulent acceleration is included, the particle distribution corresponding to $\gamma_{\max} \sim 10^4$ stretches over a wider area (see the upper middle and right plot), since macro-particles can be re-accelerated via turbulence, sustaining high-energy for a longer distance before losing a substantial portion of their energy.
In comparison to case (a), this extended spatial distribution implies a more diffuse structure of X-ray radiation attributable to IC-CMB.
In the lower panel of Fig.~\ref{fig:gam_spatial} we show the spatial distribution of the macro-particles with $\gamma_{\rm max}\sim 10^5$, responsible for the peak in the IC-CMB SED at $\sim10^{21}$\,Hz.
Similar to the former scenario, the particle distribution shows an extended morphology for case (c) as compared to other two cases, for the same reasons discussed before.
Interestingly, the spatial distributions for case (a) (left panel) and case (b) (middle panel) have a very similar structure.
The reason for this can be further investigated by comparing the $\gamma_{\max}$ histograms for case (a) and (b) (left and middle panels in Fig.~\ref{fig:histogram}), showing a similar behavior (after the peak at $\gamma_{\max}\sim 10^4$ for the latter).

In summary, we showed that, in the presence of stochastic acceleration, the emission from the radio lobe changes significantly compared to the case where STA is neglected.
With the inclusion of STA, the spatial distribution of the X-ray emitting particles through IC-CMB exhibit a wider extent (see Fig.~\ref{fig:gam_spatial}) as compared to the only DSA case, indicating a emission structure which is diffusive.

\begin{figure}
    \centering
    \includegraphics[scale=0.48]{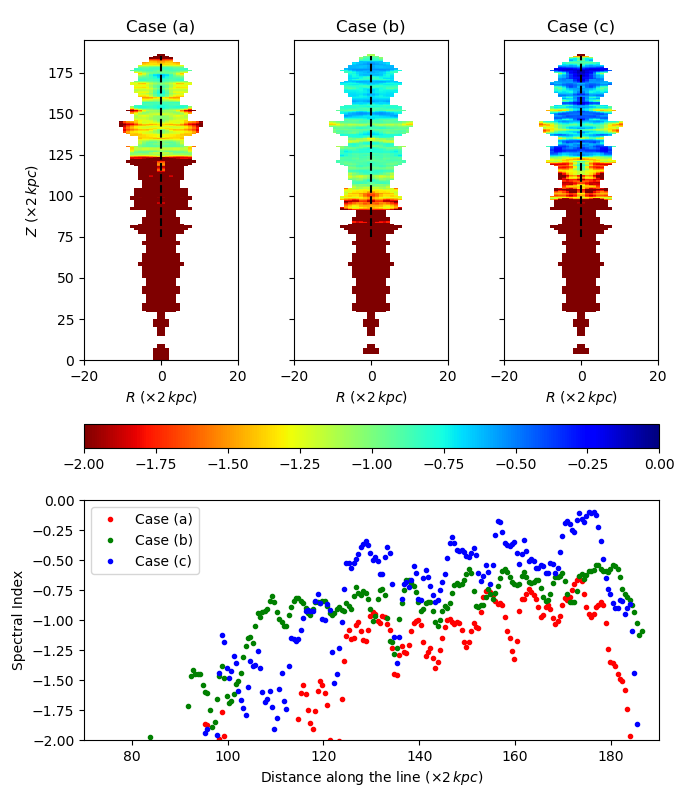}
    \caption{Spectral index map and spectral index distribution of the radio lobe for cases (a), (b) and (c). The spectral index maps are drawn considering two radio frequencies $1.5$\,GHz and $15$\,GHz. \textbf{Top left panel:} shows the spectral index map for case (a), \textbf{Top middle panel:} for case (b) and \textbf{Top right panel:} for case (c). \textbf{Bottom panel:} Spectral index distribution for all three cases.}
    \label{fig:spectral}
\end{figure}

\subsubsection{Spectral Index distribution}
In this section, we focus on the effect of STA on the radio frequency regime ($\leq 15$\,GHz). 
With the advent of several high-resolution low-frequency telescope arrays, it is possible to quantify the distribution of the spectral index in extended lobes \citep{alexander_1987,harwood_2013}.
In this regime, the emission from astrophysical systems are dominated by synchrotron radiation which follows a power-law relation with the frequency, $I_{\nu}\propto \nu^{-\delta}$, with $\delta$ being the spectral index.
In our simulation we compute the intensity from the macro-particles energy distribution (see section~\ref{sec:emiss_setup}) and further calculate the spectral index $\delta$ using the equation below:
\begin{equation}
    \delta=\frac{\log(I_{\nu_2})-\log(I_{\nu_1})}{\log(\nu_1)-\log(\nu_2)}
\end{equation}

In the top panel of Fig.~\ref{fig:spectral} we show the Gaussian-filtered spectral index maps of the radio lobe considering two frequencies, $\nu_{1}=1.5\,GHz$ and $\nu_{2}=15\,GHz$ for cases (a), (b), and (c). 
All the spectral maps show - from top to bottom - signs of spectral steepening from the outer regions of the lobe (near the bow-shock) towards the inner part.
Such a spatial distribution can be further analyzed by observing the bottom panel of the figure, where we plot the vertical distribution of the spectral index value on the path depicted by the black dashed line shown in the corresponding top panel, from the inner region of the lobe to the outer region.
The spectral index distribution behaves similarly for all three cases, showing a rapid increase followed by a softer (or almost constant) one.
By analyzing the slope of this second part, we obtain an average value for case (a) as $-1.01$, while for cases (b) and (c) it is $-0.80$ and $-0.49$ respectively.
This implies that the radiation spectrum becomes harder as one increases $\alpha$ in the lobe.
The spatial extent of the region with constant spectral index is larger for case (b) as compared to cases (a) and (c).
For case (a), this directly follows from the absence of any continuous acceleration mechanism other than shocks, and the ensuing radiative cooling of the macro-particles in the back flow over a short time scale. 
In contrast, for cases (b) and (c), STA provides additional continuous acceleration to the macro-particles.
For this reason, the macro-particles could be able to radiate for a longer amount of time and the value of the spectral index could be maintained for a longer distance. 
Additionally, due to faster turbulence decay, case (c) maintains the spectral index for a shorter spatial extent as compared to case (b). 

Our results have shown that the signature of continuous acceleration of particles due to stochastic turbulence impact on several observables, including the spectral index variation along the lobe. 
We also observed that while, with increasing $\alpha$ the spectral index value inside the lobe increases owing to the shorter acceleration timescale, the extent of the region with constant spectral index decreases due to turbulence decay.
We have discussed the implications of the synthetic measures quantified in section~\ref{sec:results} with multi-wavelength observational signatures in the next section. 

\section{Summary and Discussion}
\label{sec:summary}

In this work we have presented 2D axisymmetric large-scale numerical simulations of AGN jets using a fluid-particle hybrid approach, in order to focus on particle acceleration processes and emissions from radio lobes.
In spite of their limitations, and owing to the prohibitive cost of 3D computations, 2D models still provide fundamental insights on the interplay between different acceleration mechanisms and their influence on emission signatures.

Further owing to the multi-scale nature of the system, the underlying turbulence is considered as a sub-grid manner and its effect on the cosmic ray transport is modelled with a phenomenologically motivated ansatz for the turbulent acceleration timescale that can mimic the turbulence decay process usually observed in various astrophysical sources.
By introducing this timescale, we solve the cosmic ray transport equation to evolve their energy distribution, accounting for diffusive shock acceleration (DSA), stochastic turbulent acceleration (STA) as well as for radiative losses (synchrotron and inverse Compton), as implemented in the PLUTO code by \cite{Vaidya_2018}.
We explore different scenarios by selectively including or excluding the aforementioned acceleration mechanisms, and study their effects on the emission signatures of the radio lobes.

Below, we summarise the primary results from this work.
\begin{itemize}

\item We observe significant modification of the energy spectra of macro-particles when turbulent acceleration is included in addition to DSA as compared to only shock acceleration case.
The interplay of DSA, STA and and turbulent decay results in features such as flattening of the spectrum in the low-energy region and a dynamically evolving high energy cut-off. 
These features produce curvature in the particle spectrum which can further manifest in the emission properties of the radio lobe \citep{duffy_2012}.

\item  The analysis of the maximum attainable energy results in a unimodal PDF with a broken power-law tail when only shock acceleration is accounted for, while when both DSA and STA are included, the PDF exhibits a bimodal structure. 
Further, the PDF of the average energy ($\gamma_{\rm avg}$) for each macro-particle shows a power-law profile with an exponential cut-off on inclusion of STA.
These distributions closely resemble the case in which STA is mediated by continuous particle injection and escape \citep[see Fig.~2b of ][]{katarzynski_2006}.
Here particle injection due to shocks act as a source while the escape is due to turbulence decay. 
The lobe integrated spectrum exhibits a broken power-law like structure for DSA, whereas with STA it displays a high energy cut-off in addition to the low energy break. 
The position of the low energy break corresponds to the $\gamma$ where radiative loss time becomes equal to the dynamical time. 
The integrated spectrum generated by including STA can be utilised as a consistent input for one-zone radio lobe modelling that accounts for particle acceleration due to turbulence.

\item Further analysis of STA and its effect on sustaining the particle's energy against radiative cooling is performed through the evolution of $B_{\rm eq}/B_{\rm dyn}$ histograms, showing for all the three cases, that the system evolves to a state where $B_{\rm eq}\sim B_{\rm dyn}$.
However, with STA, the corresponding distributions become wider when compared to the only shock scenario, as a result of the additional energization.

\item The study of the synthetic SED of the simulated source demonstrates the existence of additional peaks in the radio band due to synchrotron emission and in the X-ray band through the IC-CMB mechanism when STA is taken into account.
Further analysis of the spatial distribution of the macro-particles corresponding to these additional peaks implies a more extended and diffuse emission in the X-ray band owing to the interplay of the two acceleration mechanisms.
The extent of the spatial distribution is further observed to be modulated by changing the value of $\alpha$ (see Eq.~\ref{eq:acc_time}).
This implies that, with an appropriate choice of $\alpha$, one might achieve diffuse emission around localized regions inside the radio lobe \citep[for example diffuse synchrotron emission around the hotspot of 3C445, see ][]{prieto_2002}.

\item The radio frequency spectral maps along with the spectral index profile inside the lobe indicate a harder emission spectrum due to STA as compared to the DSA case. 
The spectral index is observed to remain constant over a distance inside the radio lobe whose length gets modulated with the efficiency of the turbulent acceleration.  
The value of the spectral index in this region is $\sim -0.49$ for case (c), for case (b) it is $\sim -0.8$ and for case (a) it is $\sim -1.01$.
This kind of behaviour has also been found in various observations of radio lobes \citep{parma_1999}.
Radio lobes of parsec-scale AGN jets have been observed to exhibit similar characteristics \citep{hovatta_2014}.
However, it should be noted that from observation of radio lobes there is no evidence of spectral index $\approx-0.5$ or higher.
This, consequently, may impose a limit to the extent and the effectiveness of STA in the actual radio lobes.

\end{itemize}
\subsection*{Present Limitations and Future Extension}
The results shown in the present study represent a first step towards a more realistic description of the complex interaction between the turbulent radio lobe material and the non-thermal particles, and it is certainly limited by a number of considerations.

2D axisymmetric models, for instance, are similar to 3D models only in the case of stable jets and homogeneous media.
Time-dependent jet propagation is known to be prone to 3D instabilities (e.g., Kelvin–Helmholtz and current-driven modes) that cannot be captured by axisymmetric models \citep[see, e.g.][]{mignone_2010,bodo_2013,bodo_2016}.
These instabilities are known to have an effect on the jet emission \citep{acharya_2021,borse_2021} and can induce a range of non-axisymmetric structures, such as filaments and shocks along jets and in the back-flowing zone \citep[see for example][]{tregillis_2001,matthews_2019}.
Such non-axisymmetric structures are known to enhance the turbulence inside the back-flowing region and hence would strongly influence particle mixing \citep{jones_1999}.  

Another potential issue with 2D axisymmetric simulations is that the induction equation (Eq.~\ref{eq:induction}) does not allow, because of the $\partial_{\phi}=0$ condition, conversion of toroidal magnetic field ($B_{\phi}$) to poloidal field \citep{porth_2013}.
This leads to the continuous amplification of the injected $B_{\phi}$ component in the computational domain over time, eventually affecting the jet dynamics.
However, for this work we consider a very small $B_{\phi}$ value to lessen any substantial impact on the dynamics.
Nevertheless, 2D computations still allow our method to be tested with finer grid spacing providing better resolution across shocked structures. 
This would be computationally expensive in the fully 3D case.

Additionally, we also consider an un-magnetized ambient medium in the expectation that the magnetic field in the ambient medium will have minimal impact on the non-thermal particle transport within the lobe.

Our simulations describe the interaction between cosmic ray particles and jet materials although the former behaves essentially as a passive scalar without backreaction on the fluid.
Future extension will consider more exhaustive two-fluid approaches by also taking into account energy and momentum transfer between the two components in a self-consistent way \citep{girichidis_2020,ogrodnik_2021}.
It should be emphasized that the employment of parameters in our model is an unwanted, albeit necessary, consequence of the fact that large scale simulations cannot possibly resolve (and therefore sample) the small-scale turbulence regions.
Sub-scale micro-physical processes (such as turbulent acceleration timescale or MHD turbulence damping rate) must therefore be encoded through a sub-grid recipe.
In this work, in fact, we consider a one parameter exponentially decaying hard-sphere turbulence as a model of STA inside the radio lobe, with certain values for the parameter ($\alpha=10^{4}, 10^{5}$) and compute the emission signatures from the radio lobes via synchrotron and IC-CMB processes.

Future extensions of this work will hopefully consider fully 3D investigations, where the impact of non-axisymmetric plasma instabilities may deeply affect the morphology.
Additionally, the sub-grid prescription of turbulence decay has a crucial role in governing some of the essential properties of emission.

%
%
\bibliographystyle{mnras}
\bibliography{lobe}
\end{document}